\documentclass[debug,overfull,figures,info,cite]{epl2}

\usepackage{amsmath}
\usepackage{graphicx}
\usepackage{amssymb}
\usepackage{bm}

\newcommand{\be}{\begin{equation}}
\newcommand{\ee}{\end{equation}}
\newcommand{\bea}{\begin{eqnarray}}
\newcommand{\eea}{\end{eqnarray}}

\title{Role of symmetry in the interplay of $T=0$ quantum-phase transitions with unconventional
$T>0$ transport properties in integrable quantum lattice systems}
\author{J. M. P. Carmelo\inst{1,2}\thanks{E-mail: \email{carmelo@fisica.uminho.pt}}
\and S.-J. Gu\inst{3} \and N. M. R. Peres\inst{1}}
\shortauthor{J. M. P. Carmelo \etal}
\institute{\inst{1} GCEP-Center of Physics, U. Minho, Campus
Gualtar, P-4710-057 Braga, Portugal\\
\inst{2}Massachusetts Institute of Technology, Cambridge, MA 02139-4307, USA\\
\inst{3}Department of Physics and ITP, The Chinese University of Hong Kong,
Hong Kong, China}

\abstract{ We show that a generalized charge $SU(2)$ symmetry of the
one-dimensional (1D) Hubbard model in an infinitesimal flux $\phi$ generates
half-filling states from metallic states which lead to a finite charge
stiffness $D(T)$ at finite temperature $T$, whose $T$ dependence we study. Our
results are of general nature for many integrable quantum lattice systems,
reveal the microscopic mechanisms behind their exotic $T>0$ transport
properties and the interplay with $T=0$ quantum-phase transitions, and
contribute to the further understanding of the transport of charge in systems
of interacting ultracold fermionic atoms in 1D optical lattices, quasi-1D
compounds, and 1D nanostructures.}

\begin{document}

\maketitle

There has been a recent increasing interest in the unusual transport and spectral
properties of systems of interacting ultracold fermionic atoms in one-dimensional (1D)
optical lattices \cite{Jaksch} and a renewed interest in those of quasi-1D carbon
nanotubes, ballistic wires, and quasi-1D compounds \cite{Hiro,Sing}. Quantum effects are
strongest at low dimensionality leading to unusual phenomena such as charge-spin
separation at all energies \cite{Sing} and persistent currents in mesoscopic rings
\cite{Eckern}. The ultracold atom technology can realize a 1D closed optical lattice
experimentally with a tunable boundary phase twist capable of inducing atomic
"persistent currents" \cite{Jaksch}. Thus, the further understanding of the microscopic
mechanisms behind the transport of charge in low-dimensional  correlated systems and
materials is a topic of high scientific and technological interest. Recently there has been
an enormous grow in the literature on the quantum phase transitions in such
systems \cite{Belitz}. For instance, the relation of the unusual $T>0$ charge-transport
properties observed in low-dimensional correlated systems to their $T=0$ quantum phase
transitions is also a problem which is far of being fully understood.

For the 1D Hubbard model \cite{Lieb} or any other interacting electronic quantum
system defined in a 1D lattice of $N_a$ sites of length $L=aN_a=N_a$
($k_B =\hbar =a = -e=1$) the $T>0$ ideal
conducting and insulating behaviors are defined in terms of the real-part of the optical
conductivity for finite temperatures $T>0$, which reads
$\sigma (\omega) = 2\pi\,D\,\delta (\omega) + \sigma_{reg} (\omega)$.
Here the charge stiffness is given by $D=D(N,T)=d/Z$ where
$Z=Z (N,T)=\sum_{m}p_m$ is the partition function of the $N$-electron system,
the $m$ summation runs over the eigenstates $\vert m\rangle$ of energy
$E_m$, and $p_m=e^{-E_m/T}$ are the corresponding Boltzmann weights.
At $T=0$ the charge stiffness is related to the boundary energy of the lattice
model under twisted boundary conditions \cite{Shastry} and for $T\geq 0$
$d=d(N,T)$ can be related to the thermal average of curvatures of
energy levels subject to an infinitesimal flux $\phi$ (in units of $2\pi/L$
and $L\rightarrow\infty$),
\begin{equation}
d = {1\over L}\sum_{m}p_m\,{1\over 2}
{\partial^2 E_m (\phi)\over\partial \phi^2}\Big\vert_{\phi\rightarrow\, 0}
\, . \label{D}
\end{equation}
At $T>0$ a normal conductor is such that $D(N,T)=0$ and
$\sigma (\omega\rightarrow\,0)>0$, whereas an ideal conductor is characterized
by $D (N,T)>0$. Furthermore, an insulator can develop into a normal conductor,
become an ideal conductor, or display ideal insulating behavior such that
$D(T)=0$ and $\sigma (\omega\rightarrow\,0)=0$. 1D integrable systems
whose Hamiltonians commute with the current operator
behave as ideal conductors and ideal insulators for the densities corresponding
to the $T=0$ metallic and insulating quantum phases, respectively.
There is strong numerical evidence that the same occurs concerning the
$T>0$ ideal conducting behavior for densities referring to $T=0$ metallic phases
of 1D integrable quantum systems whose Hamiltonian does not
commute with that operator, such as the 1D Hubbard model \cite{Prelovsek,Nuno}.
However, whether for the densities corresponding to the $T=0$ insulating quantum
phases such systems behave as ideal insulators remains an open question.
In contrast, the non-integrable 1D interacting systems are generic conductors
and activated ones in the metallic and insulating phases, respectively.

In this Letter we clarify such an issue by showing that a generalized charge $SU(2)$
symmetry of the 1D Hubbard model in an infinitesimal flux $\phi$ generates half-filling
states from metallic states which lead to a finite half-filling charge stiffness
$D=D(N_a,T)$ at finite temperature $T$. We study its $T$ dependence by
characterizing its different behaviors in three well-defined temperature regimens.
Interestingly, the above half-filling states do not belong to the integrability Hilbert
subspace where the model is solvable by the Bethe ansatz \cite{Lieb}. The latter
subspace is spanned by the lowest-weight states (LWSs) of both the $\eta$-spin
and spin $SU(2)$ algebras \cite{2}. When defined in such a subspace, the quantum
problem has an infinite number of conservation laws \cite{conserva}: as in the
corresponding classical systems, the occurrence of an infinite number of such
laws is behind the integrability of the quantum models. Moreover, we present
numerical evidence that when acting onto the integrability subspace the
1D Hubbard model is an ideal insulator. Indeed, we find that in the thermodynamic
limit the half-filling states that span such a subspace do not carry charge current,
consistently with the general predictions of Ref. \cite{Prelovsek}. Importantly,
our results provide useful information about the microscopic mechanisms
behind the interplay of the $T=0$ Mott-Hubbard insulator - metal
quantum-phase transition with the $T>0$
thermal properties in 1D correlated systems and
new insights into the exotic microscopic processes of charge transport in systems
of interacting ultracold fermionic atoms in 1D optical lattices, carbon nanotubes,
ballistic wires, and quasi-1D compounds \cite{Jaksch,Hiro,Sing}.

The 1D Hubbard model in an infinitesimal flux $\phi$ reads,
\begin{equation}
\hat{H} (\phi) = -t\sum_{j,\,\sigma}[\,e^{i\phi}\,c_{j,\,\sigma}^{\dag}
c_{j+1,\,\sigma} + h. c.] + U\,{\hat{D}}_{-} \, , \label{H}
\end{equation}
where  ${\hat{D}}_{-}=\sum_{j}\{[\hat{n}_{j,\uparrow}-1/2][\hat{n}_{j,\downarrow}-1/2]+1/4\}$,
$c_{j,\,\sigma}^{\dagger}$ and $c_{j,\,\sigma}$
(and below $c_{k,\,\sigma}^{\dagger}$ and $c_{k,\,\sigma}$) are
spin-projection $\sigma =\uparrow ,\downarrow$ electron operators at site $j=1,2,...,N_a$
(of momentum $k$), the number $N_a$ is even and very large, and
$\hat{n}_{j,\,\sigma}=c_{j,\,\sigma}^{\dagger}\,c_{j,\,\sigma}$. We use periodic boundary
conditions and consider zero spin density $[n_{\uparrow}-n_{\downarrow}]=0$
and electronic densities $n=[n_{\uparrow }+n_{\downarrow}]$
and corresponding hole concentrations $\delta =1-n$ in the range
$0\leq\delta\leq 1$, where
$n_{\sigma}=N_{\sigma}/L=N_{\sigma}/N_a$ and $N_{\sigma}$ such that
$N=[N_{\uparrow}+N_{\downarrow}]$ is the number of $\sigma$ electrons.
The following $\phi =0$ properties of the Hamiltonian (\ref{H}) play
an important role in our study: (i) Its zero-energy level refers to the
$\delta =0$ insulating ground state, $E_0(N_a)=0$, and is located in
the middle of the Mott-Hubbard gap \cite{Lieb},
$\Delta_{MH}=[(4t)^2/U]\int_1^{\infty}d\omega[\sqrt{\omega^2
-1}/\sinh (2\pi t\omega/U)]$, such that
$\Delta_{MH}\approx [8\sqrt{Ut}/\pi]e^{-2\pi t/U}$ for $U/t<<1$ and
$\Delta_{MH}\approx U-4t$ for $U/t>>1$; (ii) The ground-state energies
associated with even values of $N\leq N_a$ are such that $E_0 (N-2)>E_0 (N)$
where $E_0 (N)$ changes from $[4N_a t/\pi][1-\cos (\pi\delta/2)]$ for $U/t<<1$ to
$[\delta N_aU/2]\{1-[4t\sin (\pi\delta)/\pi\delta U]\}$ for $U/t>>1$; (iii)
The $T=0$ chemical potential $2\mu (N,0)$ is an increasing
function of $\delta$, changing from $4t\sin (\pi\delta/2)$ for $U/t<<1$
to $U-4t\cos (\pi\delta)$ for $U/t>>1$ and is given by
$2\mu (N,0)\approx \Delta_{MH}+\delta^2\pi^2/m^*_h$ and
$2\mu (0,0)=U+4t>\Delta_{MH}$ for $\delta$ small and $\delta =1$,
respectively, and all values of $U/t$; (iv)
$m^*_h$ is a charge mass given in Eq. (48) of Ref. \cite{91}
whose limiting values are $m^*_h\rightarrow 0$ as
$U/t\rightarrow 0$ and $m^*_h\rightarrow 1/2t$ as $U/t\rightarrow\infty$.
\begin{figure}
\includegraphics[width=8.5cm,height=4cm]{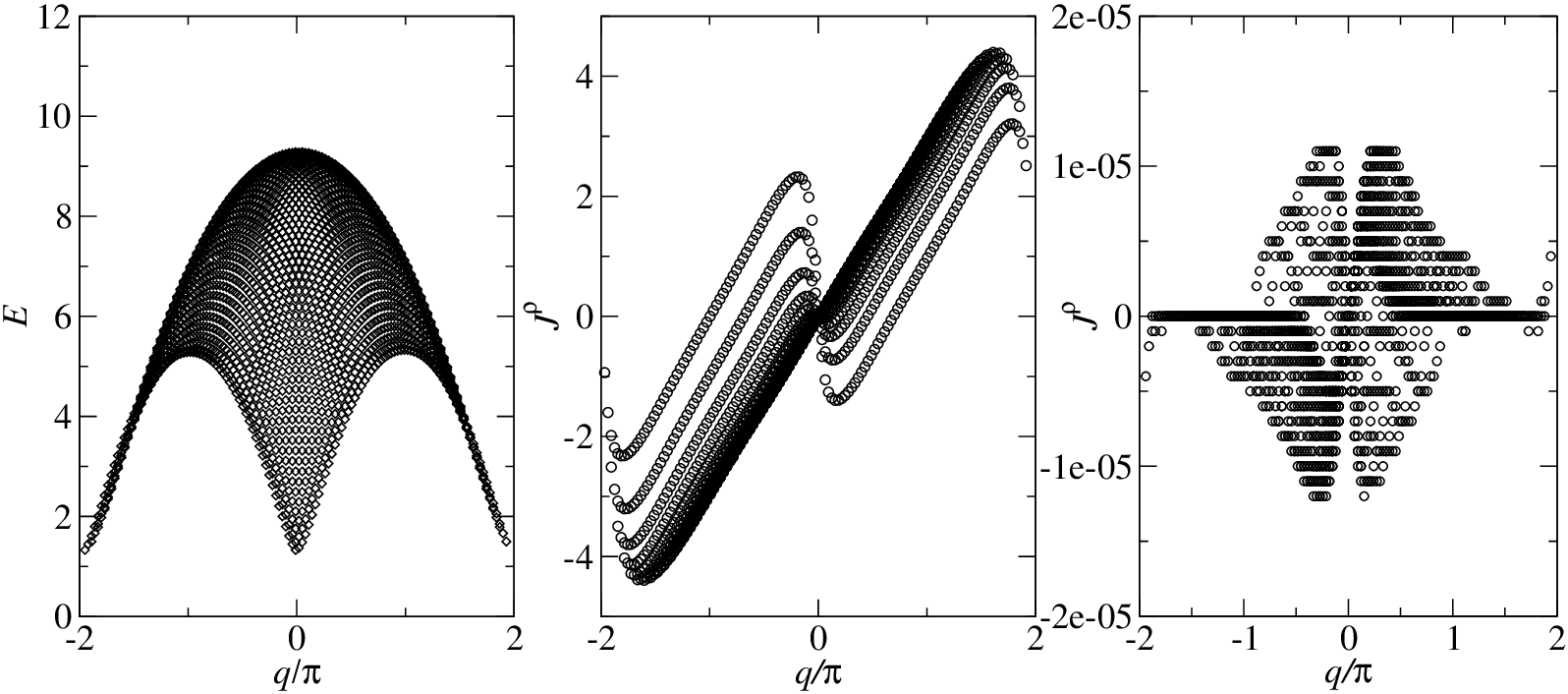}
\caption{\label{fig1} The degenerated energy spectrum and charge current spectrum in
units of $t$ of the half-filling $\eta$-spin-triplet states (left current spectrum) and
corresponding singlet states (right current spectrum) for $U/4t=1$ and $N=N_a=66$.}
\end{figure}

For $\phi =0$ the Hamiltonian (\ref{H}) commutes with the three generators of the
$\eta$-spin $SU(2)$ algebra \cite{2}. While for $\phi>0$ it does not commute with
the two off-diagonal generators of that algebra, a result which plays a central role
in our study is that for $\phi >0$ it commutes with the three generators
${\hat{\eta}}^z_{\phi}=-{1\over 2}\sum_{j}
[1-\hat{n}_{j,\uparrow}-\hat{n}_{j,\downarrow}]$, ${\hat{\eta}}^{\dagger}_{\phi}=\sum_{k}
\,c_{\pi -\phi -k ,\,\downarrow}^{\dagger}\,c_{k,\,\uparrow}^{\dagger}\, =\sum_j\,e^{i j(\pi-\phi)}c_{j,
\,\downarrow}^\dagger\,c_{j,\,\uparrow}^\dagger$, and
${\hat{\eta}}_{\phi} = \sum_{k}\,c_{k,\,\uparrow}\,c_{\pi -\phi-k,\,\downarrow}\,
=\sum_j\,e^{-i j(\pi-\phi)}c_{j,\,\uparrow}\,c_{j, \,\downarrow}$ of a {\it generalized}
charge $SU(2)$ algebra. Since for $\phi >0$ the number of electrons
$N$ and $\eta$-spin value $\eta$ remain good
quantum numbers, the diagonal generator ${\hat{\eta}}^z_{\phi}$ is independent of $\phi$.
Furthermore, note that as $\phi\rightarrow 0$ the above off-diagonal generators
of the generalized algebra become those of the $\eta$-spin algebra
\cite{2}. For $\phi >0$ the Hamiltonian (\ref{H}) remains integrable. However, as for
$\phi =0$, the generalized Bethe-ansatz equations given in Eq. (10)-(12) of Ref.
\cite{Nuno} for $\phi >0$ refer to the Hilbert subspace spanned by the LWSs of both the
$\eta$-spin and spin algebras such that $2\eta = [N_a -N]$ and $2S
=[N_{\uparrow}-N_{\downarrow}]$, where $S$ is the spin. Thus, for half filling such an
integrability subspace is spanned by the $\eta =0$, $\eta$-spin singlet states. The full
Hilbert space contains another subspace which is spanned by energy eigenstates generated
from application onto the LWSs of the off-diagonal generators of the generalized
algebras. Since here we are mostly interested in the transport of charge, for simplicity
we limit our study to the charge and $\eta$-spin degrees of freedom, yet it is
straightforward to extend our analysis to the spin degrees of freedom. Let $\vert
m,N,\phi\rangle$ be a $N$-electron energy eigenstate of the Hamiltonian (\ref{H}) for
$\phi >0$, such that $2\eta > [N_a -N]$. This state does not belong to the subspace where
the model is integrable but can be generated from the corresponding $N'$-electron LWS
such that $N' =[N_a -2\eta ]< N$ as follows,
\begin{equation}
\vert m, N,\phi\rangle = {1\over\sqrt{C}}
\left[{\hat{\eta}}^{\dagger}_{\phi}\right]^{\eta -{1\over 2}[N_a -N]}\vert m', N',
\phi\rangle\, , \label{state}
\end{equation}
where the normalization constant reads
$C = \delta_{\eta,\,{1\over 2}[N_a -N]} + \prod_{l=1}^{\eta -{1\over 2}[N_a -N]} l[2\eta +
1 - l]$. Since the Hamiltonian (\ref{H}) commutes with the generators
${\hat{\eta}}^{\dagger}_{\phi}$ and ${\hat{\eta}}_{\phi}$, the energies
$E_m (\phi) $ and $E_{m'} (\phi)$ of the above states $\vert m,N,\phi\rangle$ and
$\vert m',N',\phi\rangle$ are identical.

The microscopic mechanism described in the following implies
that, in spite of the model integrability, for finite values of $U/t$ the
$\delta=0$ charge stiffness is finite for finite $T$ values and
vanishes both as $T\rightarrow 0$ and $T\rightarrow\infty$.
\begin{figure}
\includegraphics[width=8.5cm,height=4cm]{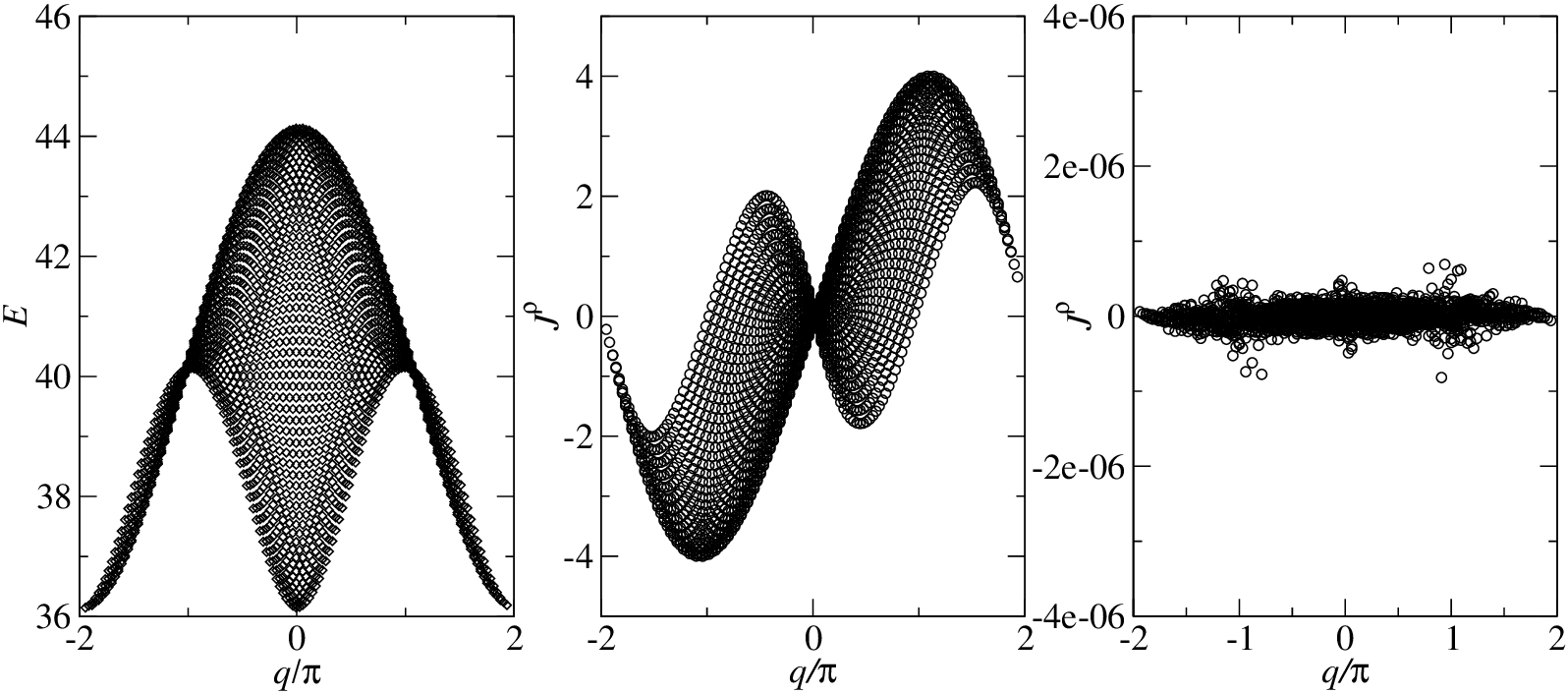}
\caption{\label{fig2} The degenerated energy spectrum and charge current spectrum in
units of $t$ of the half-filling $\eta$-spin-triplet states (left current spectrum) and
corresponding singlet states (right current spectrum) for $U/4t=10$ and $N=N_a=66$.}
\end{figure}
For each energy eigenstate $\vert m,N,\phi\rangle$ of electronic-number
$N$ and $2\eta > [N_a-N]$ there is one and only one $\eta$-spin LWS
$\vert m',N',\phi\rangle$ with the same $\eta$ value, $N'<N$, and
$2\eta = [N_a - N']$ whose energy is such that $E_{m'} (\phi)=E_{m} (\phi)$,
where $E_{m}(\phi)$ is that of $\vert m,N,\phi\rangle$.
In turn, there are $2\eta$ states $\vert
m,N,\phi\rangle$ such that $N=N_a -2\eta + 2j$ with
$j=1,2,...,2\eta$ corresponding to the same $\eta$-spin LWS $\vert
m',N',\phi\rangle$. Thus, $Z(N,T)=\delta_{N,N_a}+Z_{LWS}(N,T)+Z_{R}(N,T)
=\delta_{N,N_a}+\sum_{N'=0}^{N}Z_{LWS}(N',T)$
(and $d(N,T)=d_{LWS}(N,T)+d_{R}(N,T)=\sum_{N'=0}^{N}
d_{LWS}(N',T)$), where $\delta_{N,N_a}+Z_{LWS}(N,T)$
(and $d_{LWS}(N,T)$) corresponds to the contributions from
all the $\eta$-spin LWSs of electronic number $N$ and
$Z_{R} (N,T)=\sum_{N'=0}^{N-2}Z_{LWS}(N',T)$
(and $d_{R} (N,T)=\sum_{N'=0}^{N-2}d_{LWS}(N',T)$) is
such that the $N'$ summation refers to all LWSs with
even electronic numbers $N'=0,2,4,...,N-2$ from which one can
generate all non-LWSs with $N$ electrons. (The
above contribution $\delta_{N,N_a}$ refers to the
insulating ground state.) Such results imply that
${\rm max}\,d (N,T)=d (N_a,T)$ and thus if $d(N,T)>0$ for
$N$ even, $N<N_a$, and $T>0$ then $d (N,T)>0$ for
$N=N_a$ and $T>0$.

Without the construction of the states (\ref{state}) for $\phi >0$
the above results could not be reached. Their use reveals that
a parameter which plays key role in our study is,
\begin{equation}
\alpha (\Delta N,T)=
\bar{Z}(\Delta N,T)/Z (N_a -\Delta N,T) \, ;
\hspace{0.25cm} \bar{Z}(\Delta N,T)=
\sum_{N'=N_a -\Delta N-2}^{N_a}Z_{LWS}(N',T) \, ,
\label{alpha}
\end{equation}
where $Z (N_a -\Delta N,T)$ is the partition function
for $N=N_a -\Delta N$ and the $N'$ sum runs over
even values only. For finite values of $T$
and $U/t$ one has that $\alpha (\Delta N,T)<1$ for
$\Delta N\leq\Delta N_0 =\Delta N_0 (T)$,
where the critical value
$\Delta N_0 (T)=2,4,6,...$ increases with $T$, being such that
$\Delta N_0 (T)\rightarrow 2$ as $T\rightarrow 0$ and
$\Delta N_0 (T)\rightarrow N_a$ as $T\rightarrow\infty$.
Moreover, ${\rm min}\{\alpha (\Delta N,T)\}=\alpha (2,T)$,
${\rm max}\{\alpha (2,T)\}=\alpha (2,0)=1$, and $\alpha (2,T)<1$
for $T>0$.

Three typical regimens correspond to
(a) $2\mu (N_a,T)/T\rightarrow\infty$ when
$T<<2\mu (N_a,0)=\Delta_{MH}$,
(b) $2\mu (N_a,T)/T\approx 0$ for intermediate temperatures
$T\approx 2\mu (N_a,0)=\Delta_{MH}$, and (c)
$2\mu (0,T)/T\rightarrow -\infty$ when
$T\rightarrow\infty$. We find from the above results that for $U/t>0$ the
half-filling charge stiffness is of the general form,
\begin{equation}
D (N_a,T) = {D (N_a -\Delta N_0,T)\over 1+
F (\Delta N_0,T)}
+ \gamma (\Delta N_0,T) \, , \label{DNaT}
\end{equation}
where $F (\Delta N,T)=
\alpha (\Delta N,T)+1/Z (N_a -\Delta N,T)$
and
\begin{equation}
\gamma (\Delta N,T)= [d (N_a,T)-d (N_a -\Delta N,T)]/Z (N_a,T) \, ,
\label{ga}
\end{equation}
is such that $\gamma (\Delta N,T)\geq 0$ for all
$\Delta N=2,4,...,N_a$. For the regimen (a), $e^{-\Delta_{MH}/T}
\approx e^{-\infty}$ and thus also $e^{-2\mu (N',0)/T}
\approx e^{-\infty}$, because $2\mu (N',0)>\Delta_{MH}$
for $0\leq N'<N_a$. Thus, one has
both that ,
\begin{equation}
{Z_{LWS}(N',T)\over Z_{LWS}(N'-2,T)} \approx e^{-2\mu (N',0)/T}
\approx e^{-\infty} \hspace{0.2cm}{\rm and}\hspace{0.2cm}
{d_{LWS}(N',T)\over d_{LWS}(N'-2,T)}\approx e^{-2\mu (N',0)/T}
\approx e^{-\infty}\, , \label{ratios}
\end{equation}
and only the term $d_{LWS}(N_a-2,T)$
contributes to $d_{R} (N_a,T)=\sum_{N'=0}^{N_a-2}d_{LWS}(N',T)$.
We then find from use of Eq. (\ref{DNaT}) that,
\begin{equation}
D (N_a,T)\propto e^{-\Delta_{MH}/T}\sqrt{T/m^*_h} \, ; \hspace{0.25cm}
T<< \Delta_{MH} \, .
\label{D-T-small}
\end{equation}
In turn, as the temperature increases
and reaches the regimen (b) such that $T\approx \Delta_{MH}$, one finds
that $e^{2\mu (N',T)/T}\approx 1$ for an increasing number
of $Z_{LWS}(N',T)$ and $d_{LWS}(N',T)$ terms of
the expressions $Z_{R} (N_a,T)=\sum_{N'=0}^{N_a-2}Z_{LWS}(N',T)$
and $d_{R} (N_a,T)=\sum_{N'=0}^{N_a-2}d_{LWS}(N',T)$,
respectively, so that $Z_{R} (N_a,T)$ and $d_{R} (N_a,T)$
have significant contributions from a maximum number
of such terms. Moreover, $\delta_0 = \Delta N_0 (T)/N_a$ is
now finite and $e^{2\mu (N_a-\Delta N_0,T)/T}\approx 1$
implies that $Z (N_a -\Delta N_0,T)>>1$ and thus
$F (\Delta N_0,T)<1$ or $F (\Delta N_0,T)\approx 1$.
Therefore, since for $\delta_0$ finite the system is an ideal
conductor for $N=N_a-\Delta N_0$ \cite{Prelovsek,Nuno},
$D(N_a,T)$, Eq. (\ref{DNaT}), is finite for the regimen (b). Furthermore, the
achieved maximum contribution from the whole set of above
terms leads to a maximum value of $D(N_a,T)$ for $T\approx \Delta_{MH}$.
For increasing values of $T$, $D(N_a,T)$ becomes a decreasing function
of $T$ and finally when $T\rightarrow\infty$ the regimen (c)
such that the two ratios of Eq. (\ref{ratios}) reach the values
$e^{-2\mu (N',T)/T}\approx e^{\infty}$ and $e^{2\mu (N',T)/T}
\approx e^{\infty}$, respectively. We then find from use of
Eq. (\ref{DNaT}) that,
\begin{equation}
D (N_a,T)\propto 1/T \, ; \hspace{0.25cm}
T\rightarrow\infty \, ,
\label{D-T-large}
\end{equation}
which is the expected behavior of systems on a lattice. The overall
charge-stiffness $T$ dependence is thus similar to that of
the $\delta=0$ dimerized model plotted in Fig. 6 of Ref. \cite{Nuno},
with the dimerized gap $\Delta$ replaced by $\Delta_{MH}$.
\begin{figure}
\includegraphics[width=8.5cm,height=4cm]{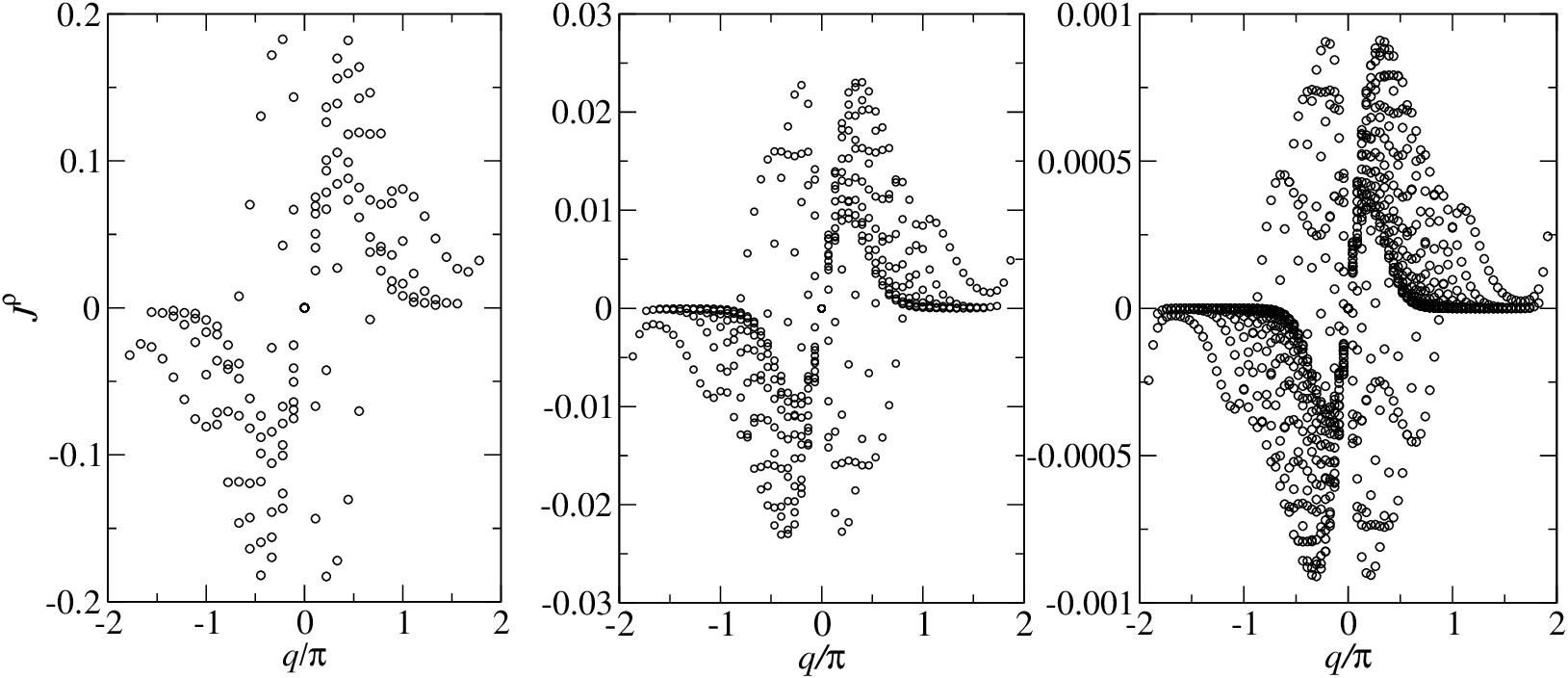}
\caption{\label{fig3} The charge current spectra in units of $t$ of the half-filling
$\eta$-spin-singlet states of Figs. 1 and 2 for $U/4t=1$ and $N=N_a=18$, $N=N_a=30$, and
$N=N_a=46$.}
\end{figure}
The temperature $T\approx \Delta_{MH}$ associated
with the maximum value of $D(N_a,T)$ is an increasing
function of $U/t$. For $U/t\rightarrow\infty$ one has that
$T\approx \Delta_{MH}\approx U-4t\rightarrow\infty$ and
thus $D (N_a,T)$ is given by Eq. (\ref{D-T-small}) for all ranges of finite
$T$ values. Thus, for $U/t\rightarrow\infty$ the regimens (b) and (c) are never
reached for finite $T$ values. In the opposite limit, when upon decreasing
the value of $U/t$ one reaches $T\approx \Delta_{MH}\approx
[8\sqrt{Ut}/\pi]e^{-2\pi t/U}$, the temperature range above $T=0$
where $D (N_a,T)$ is given by Eq. (\ref{D-T-small})
becomes smaller and smaller, until it shrinks as both
$\Delta_{MH}\rightarrow 0$ and $m^*_h\rightarrow 0$
and the limit $U/t=0$ is reached. In that limit a maximum
value $D(N_a,0)=2t/\pi$ occurs for $T=\Delta_{MH}\rightarrow 0$
and thus $D(N_a,T)$ becomes finite and largest at $T=0$.
Thus, the regimen (a) is absent for $U/t\rightarrow 0$, which implies a metallic ground
state for $U/t=0$ and $\delta =0$ and ideal conducting behavior
associated with the regimen (b) for finite temperatures. Our above results
then reveal the microscopic processes behind the $T=0$ half-filling
$U/t\rightarrow 0$ Mott-Hubbard insulator - metal phase
transition. Indeed, for finite $U/t$ values $D(N_a,T)$ displays
a maximum value at $T\approx \Delta_{MH}$ which corresponds
to the merging of the regimen (b) and vanishes both for
$T\rightarrow 0$ and $T\rightarrow\infty$ as given
in Eqs. (\ref{D-T-small}) and (\ref{D-T-large}), respectively. However, for
$U/t\rightarrow 0$ the maximum value is shifted to
$T=0$ as $\Delta_{MH}\rightarrow 0$, which is
behind the $T=0$ quantum phase transition.
The above mechanism reveals that the finite
values $D(N_a,T)>0$ for $T>0$ and $U/t>0$
result from half-filling states generated by the generalized
charge $SU(2)$ algebra from metallic states.

For $L\rightarrow\infty$ the charge stiffness can be expressed as
$D=[1/2TZL]\sum_{m}p_m\,(j_m)^2$ in terms of
the expectation values $j_m =\langle m\vert\hat{j}\vert m\rangle =
-[\partial E_m (\phi)/\partial \phi]\vert_{\phi\rightarrow\, 0}$
of the current operator $\hat{j}$ and thus $D=D (N,T)$
vanishes if $j_m=0$ for all energy eigenstates \cite{Nuno} .
In order to confirm that the $\delta =0$ finite charge currents
$j_m$ are carried by states generated from metallic states, we
provide numerical evidence that the $\delta =0$ $\eta$-spin
singlet energy eigenstates which span the integrability subspace
carry no charge current and thus that $d_{LWS}(N_a,T)=0$. Our numerical
results also confirm that the $\delta =0$, $\eta >0$ states carry
finite current, in agreement with our above analysis.
We have used the $L>>1$ Bethe-ansatz equations of Ref.
\cite{Nuno} with $L$ finite and checked that our finite-size results
are in excellent numerical agreement with all known quantities.
In Figs. 1 and 2 we plot the $\delta =0$ energy spectrum as a function
of the momentum of the $\eta =1$, $\eta$-spin-triplet energy
eigenstates and of the corresponding $\eta =0$, $\eta$-spin-singlet
states whose energy spectrum is degenerated with that of the former
states for two values of $U/4t$. Such $\eta =0$ states have one
charge string of length one \cite{Nuno}. These
figures also display the charge-current spectrum of these two types
of states. Note that the current value is finite for the $\eta$-spin
triplet states, in agreement with our above analysis, whereas for the
$\eta =0$ states it is very small for the finite system and vanishes
as $L\rightarrow\infty$. In figure 3
we show the current spectrum of the same $\eta =0$ states for
$U/4t=1$ and different values of $N_a$. The data of the figure confirm
that the current values decrease for increasing values of $L$ and
our extrapolation for $L\rightarrow\infty$ reveals that they vanish in that limit.
\begin{figure}
\includegraphics[width=8.5cm,height=4cm]{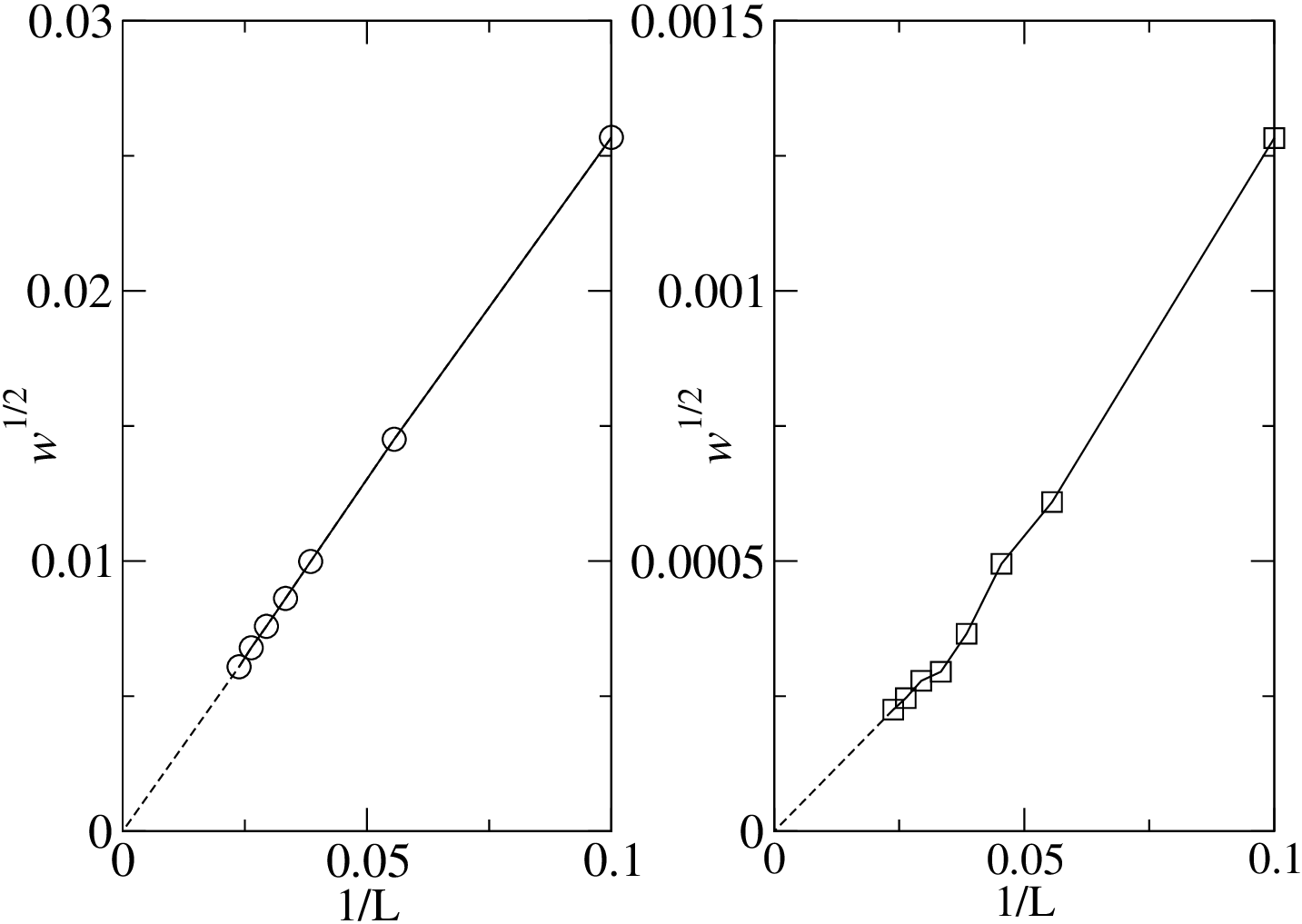}
\caption{\label{fig4} The square root of the summation of the
currents of the half-filling $\eta$-spin-singlet energy
eigenstates with one charge string of length two (left) and two
strings of length one (right) divided by the number of these
states for $U/4t=10$ as a function of $1/L$. The dashed lines
correspond to the expected behavior for $1/L\rightarrow 0$.}
\end{figure}
Finally, we study the current expectation values of all $\delta =0$, $\eta =2$
states and of the corresponding $\eta$-spin-singlet states whose energy spectrum is
degenerated with that of these states. There are two types of such $\eta =0$ states,
which have one charge string of length two and two charge strings of length
one, respectively \cite{Nuno}. We consider the square root of the summation of the currents of
these two classes of states. The results are plotted in Fig. 4 for $U/4t=10$ as a function of
$1/L$ and indicate that the currents of these states also vanish as $1/L\rightarrow 0$. A
similar analysis for other values of $U/4t$ and random samples of $\eta$-spin-singlet states
whose energy is degenerated with that of $\delta =0$, $\eta >2$ states also led to
vanishing currents as $L\rightarrow\infty$.

In this Letter we have shown that at $N=N_a$ the 1D Hubbard model is
not an ideal insulator because a generalized charge $SU(2)$
symmetry of the model in an infinitesimal flux
generates half-filling states from metallic states that carry finite current.
Moreover, we found evidence that when acting onto the
Hilbert subspace spanned by the energy eigenstates associated
with integrability the 1D Hubbard model is at half filling an ideal
insulator. Our results also reveal the microscopic
mechanisms behind the interplay of the half-filling $T=0$ Mott-Hubbard
insulator - metal quantum-phase transition with the model unusual
$T>0$ properties and contribute to the further understanding of the
transport of charge in systems of interacting ultracold fermionic atoms
in 1D optical lattices, quasi-1D compounds, and 1D nanostructures.

We thank P. A. Lee for discussions and the support of the Gulbenkian Foundation,
Fulbright Commission, ESF Science Program INSTANS, and grants POCTI/FIS/58133/2004 and
RGC CUHK 401504.

\end{document}